\documentstyle[12pt,epsf]{article}

\def\Tr{\mbox{Tr}\,}

\def\ds{\displaystyle}

\makeatletter

\def\lesssim{\mathrel{\mathpalette\vereq<}}
\def\vereq#1#2{\lower3pt\vbox{\baselineskip1.5pt \lineskip1.5pt
\ialign{$\m@th#1\hfill##\hfil$\crcr#2\crcr\sim\crcr}}}
\makeatother

\begin{document}

\noindent ULM--TP/97-7 \\
August 1997
\vspace{3.0cm}

\centerline{\LARGE Uniform Approximation for}
\vspace{0.3cm}
\centerline{\LARGE Period-Quadrupling Bifurcations}
\vspace{2.0cm}
\centerline{\large Martin Sieber$^1$ and Henning Schomerus$^2$}
\vspace{0.5cm}

\noindent $^1$ Abteilung Theoretische Physik,
Universit\"at Ulm, D-89069 Ulm, Germany\\
\noindent $^2$ Fachbereich Physik, Universit\"at-Gesamthochschule
Essen, D-45117 Essen, Germany

\vspace{2.0cm}
\centerline{\bf Abstract}
\vspace{0.5cm}

We derive a uniform approximation for semiclassical contributions
of periodic orbits to the spectral density which is valid
for generic period-quadrupling bifurcations in systems
with a mixed phase space. These bifurcations involve three
periodic orbits which coalesce at the bifurcation. In the
vicinity of the bifurcation the three orbits give a 
collective contribution to the spectral density while the
individual contributions of Gutzwiller's type would diverge
at the bifurcation. The uniform approximation is obtained by
mapping the action function onto the normal form
corresponding to the bifurcation. The present article is a
continuation of previous work in which uniform approximations
for generic period-$m$-tupling bifurcations with $m \neq 4$
were derived.

\vspace{2.5cm}

\noindent PACS numbers: \\
\noindent 03.20.+i ~ Classical mechanics of discrete systems: 
general mathematical aspects. \\
\noindent 03.65.Sq ~ Semiclassical theories and applications. \\
\noindent 05.45.+b ~ Theory and models of chaotic systems.

\vspace{0.3cm}
\noindent{\it Submitted to Journal of Physics A}

\newpage

\section{Introduction}
\label{secintro}

Semiclassical approximations for the density of states of
a quantum system can be expressed in terms of classical
periodic orbits \cite{Gut71,BB72,BT76,BT77a,Gut90}.
The form in which the periodic orbits contribute
in these approximations is not unique; it rather depends
on the characteristics of the classical motion.
Up to now, complete approximations in terms of periodic
orbits have been derived only for the cases of either integrable
or globally chaotic classical motion.
In the more general situation of a mixed phase space
difficulties arise due to bifurcations, i.\,e.\ to
the coalescence of two or more periodic orbits as
the energy or some external parameter is varied.
In the present article we describe the treatment
of a generic type of bifurcation. We derive a uniform
approximation for the joint contribution of orbits
that participate in a period-quadrupling bifurcation.

One way to derive semiclassical approximations
is by starting from Feynman's path integral
and evaluating all integrals semiclassically.
The periodic-orbit contributions then arise
from the stationary points of oscillatory integrals.
If the periodic orbits are isolated the
integrals can be evaluated by a stationary-phase
approximation and one obtains Gutzwiller's
trace formula. At bifurcations of periodic orbits,
however, different stationary points coalesce
and the stationary-phase approximation breaks down.
This is because the different stationary points
cannot be treated separately in the vicinity of
the bifurcation. Instead one has to treat them
collectively. This is done
in terms of certain canonical integrals with the
same structure of stationary points.

The characteristic arrangement of stationary
points in the vicinity of a bifurcation is
described by its normal form. The generic normal
forms for bifurcations in autonomous
systems with two degrees of freedom or,
equivalently, two-dimensional
area-preserving maps have been classified
by Meyer and Bruno \cite{Mey70,Brj70,Bru72}.
They depend on the ratio $m$ of the primitive
periods of the periodic orbits which coalesce
at the bifurcation. The corresponding bifurcations
are named period-$m$-tupling bifurcations. Ozorio
de Almeida and Hannay derived transitional (or local)
approximations for the contributions of
periodic orbits near generic bifurcations
\cite{OH87}. These approximations are expressed
in terms of canonical catastrophe
diffraction integrals. They are transitional
approximations because they are valid in the vicinity
of the bifurcation. In farther distance to
the bifurcation, however, they do not yield
the correct amplitudes in Gutzwiller's
approximation for the contribution of isolated
periodic orbits.

In two previous articles we extended the
results of Ozorio de Almeida and Hannay by
deriving uniform approximations which interpolate
between the transitional approximation at the bifurcation
and Gutzwiller's approximation for isolated periodic
orbits \cite{Sie96,SS97} (see also \cite{Sie97}).
These uniform approximations were obtained by including
higher-order corrections to the normal form expansion
and then simplifying the integrals by appropriate 
coordinate transformations. The derivations were
done for the cases $m > 4$ and $m < 4$, respectively.
In the present paper we treat the remaining case $m=4$.
This case is more complicated than the
other cases since it involves three periodic orbits
whose action differences all increase with the 
same power of the parameter that describes the
distance to the bifurcation.
We use a different method for the derivation
of the uniform approximation than previously:
We apply techniques of catastrophe theory for obtaining
uniform approximations for oscillatory integrals with
almost coincident stationary points by performing a
mapping onto the normal form \cite{Dui74,Con76,Ber76}.
The motivation is to provide a more elegant derivation;
the results are, of course, the same for both approaches.

In the following section we present the
uniform approximation that is derived in appendix \ref{seca1} and
discuss several limiting cases. The result
is given for autonomous systems with two degrees of freedom and for
two-dimensional area-preserving maps. We apply the 
uniform approximation numerically to the kicked top
and discuss the limits of its validity.

\section{The uniform approximation}
\label{secdofe}

The semiclassical contributions of periodic orbits to the
spectral density can be obtained by expressing the density
in terms of the trace of the (retarded) Green function,
\begin{equation} \label{sec2a}
d(E) = \sum_n \delta(E - E_n) 
     = - \frac{1}{\pi} \mbox{\bf Im\,} \Tr G(E) \; ,
\end{equation}
and evaluating the trace semiclassically in the vicinity of the
orbits. If the Green function is expressed in a mixed
coordinate-momentum representation and the integrals over
the components of the coordinates along the periodic orbits
are carried out one arrives at an integral expression
of the form 
\begin{equation} \label{sec2b}
d_\xi(E) \approx \frac{1}{2 \pi^2 \hbar^2} \, \mbox{\bf Re} 
\int_{-\infty}^\infty \! \mbox{d}q' \, 
\int_{-\infty}^\infty \! \mbox{d}p  \, \frac{1}{r} \,
\frac{\partial \hat{S}}{\partial E} \,
\left| \frac{\partial^2 \hat{S}}{\partial p \partial q'}
\right|^{\frac{1}{2}}
\exp \left[ \frac{i}{\hbar} \hat{S}(q', p, E) 
- \frac{i}{\hbar} q' p - \frac{i \pi}{2} \nu \right] \; .
\end{equation}
For a more detailed derivation of this integral see
\cite{Sie96,SS97}. In (\ref{sec2b}) the origin of
the coordinate system is located on a central periodic
orbit with repetition number $r$. Furthermore, $p$ and $q'$ are
coordinates in a Poincar\'e surface of section
perpendicular to the orbit, and $\hat{S}(q', p, E)$
is the generating function for the $r$-th iterate of
the Poincar\'e map. It obeys the conditions
\begin{equation} \label{sec2bb}
\frac{\partial \hat{S}}{\partial q'} = p'
\; , \; \; \;
\frac{\partial \hat{S}}{\partial p}  = q
\; , \; \; \;
\frac{\partial \hat{S}}{\partial E} = T \; ,
\end{equation}
where the primed quantities are the final coordinates,
the unprimed quantities the initial coordinates, and
$T$ is the time from initial to final point. The index
$\xi$ of the spectral density denotes the contributions
from a group of orbits in the vicinity of the central
periodic orbit. The periodic orbits are the solutions
of
\begin{equation}\label{sec2bc}
\frac{\partial \hat{S}}{\partial q'} = p
\; , \; \; \;
\frac{\partial \hat{S}}{\partial p}  = q' \; ,
\end{equation}
and correspond to stationary points of the integral (\ref{sec2b}).

If the integral in (\ref{sec2b}) is evaluated in
stationary-phase approximation one obtains Gutz\-willer's
contributions of isolated periodic
orbits. For a periodic orbit labelled by $\gamma$ this
contribution is given by
\begin{equation} \label{sec2c}
d_\gamma(E) = \frac{A_\gamma(E)}{\pi \hbar} \cos \left(
\frac{S_\gamma(E)}{\hbar} - \frac{\pi}{2} \nu_\gamma \right) \; ,
\end{equation}
where
\begin{equation} \label{sec2d}
A_\gamma(E) = \frac{T_\gamma(E)}{r_\gamma
\sqrt{|\Tr M_\gamma - 2|}} \; .
\end{equation}
Here $S_\gamma$, $T_\gamma$, $r_\gamma$, $M_\gamma$ and
$\nu_\gamma$ are, respectively, the action, period, repetition number,
stability matrix, and Maslov index of the orbit.

In the vicinity of a bifurcation a stationary-phase
evaluation of the integrals in (\ref{sec2b}) is not
appropriate. Instead one has to integrate collectively over all
stationary points which are involved in the bifurcation.
This is done by inserting the normal form of the
generating function $\hat{S}(q', p, E)$ for the
considered bifurcation into (\ref{sec2b}). For a
generic period-quadrupling bifurcation the repetition
number $r$ is a multiple of 4; we denote in
the following $l=r/4$. The normal form for this case is
given by
\begin{eqnarray} \label{sec2e}
\hat{S}(q', p, E) &=&
S_0(E) + q' p - \frac{\varepsilon}{2}(q'^2 + p^2)
- \frac{a}{4} (q'^4 + 2 p^2 q'^2 + p^4) 
- \frac{b}{4} (q'^4 - 6 p^2 q'^2 + p^4)
\nonumber
\\
&=&
S_0(E) + q' p 
- \varepsilon I - a I^2  - b I^2 \cos(4 \Phi) \; ,
\end{eqnarray}
where $p = \sqrt{2 I} \cos \Phi$ and $q' = \sqrt{2 I}
\sin \Phi$. The parameter $\varepsilon$ is zero
at the bifurcation. The normal form (\ref{sec2e}) is
obtained from an expansion of the Hamiltonian in the
vicinity of the central periodic orbit, and $S_0$ is
the action of this orbit.

The generic period-quadrupling bifurcation that is
described by the normal form (\ref{sec2e}) involves
three periodic orbits, a central orbit and two
satellite orbits. The bifurcation occurs in two
different forms depending on the relative
magnitude of the two coefficients $a$ and $b$ in
(\ref{sec2e}). In the case $|a| < |b|$ 
there are two real orbits and one complex
orbit before and after the bifurcation, the stable
central orbit, an unstable satellite orbit and
a complex satellite orbit. At the bifurcation
one of the satellites becomes complex and the
other one becomes real. For $|a| > |b|$ the
two satellite orbits are both complex on one
side of the bifurcation and both real on the
other side of the bifurcation, where one of them
is stable and the other is unstable. The central
orbit is real and stable on both sides of the
bifurcation.

In the vicinity of the bifurcation, i.\,e.\ for sufficiently
small values of $\varepsilon$, the contribution of the orbits
can be described by the transitional approximation of Ozorio
de Almeida and Hannay. It is obtained by approximating the
pre-exponential factor in (\ref{sec2b}) by its value at the
origin and evaluating the integral with the normal form
(\ref{sec2e}) for the action. This yields the semiclassical
contribution in terms of the diffraction catastrophe integral
for the catastrophe $X_9$. In farther distance from the
bifurcation the transitional
approximation splits up into a sum of separate contributions
of Gutzwiller's type. However, in this
limit the semiclassical amplitudes come out with fixed relationships
which are, in general, not in accordance with the  periods and
stabilities of the orbits.
In more detail, the approximation is good as long as the following
relations between the monodromy matrices and the periods of
the orbits hold,
\begin{eqnarray} \label{sec2f}
4 \Delta S_{21} \, \mbox{Tr\,} M_0 
+ \Delta S_{20} \, \mbox{Tr\,} M_1 &=& 8 \Delta S_{21} + 2 \Delta S_{20} 
\nonumber \\
4 \Delta S_{12} \, \mbox{Tr\,} M_0 
+ \Delta S_{10} \, \mbox{Tr\,} M_2 &=& 8 \Delta S_{12} + 2 \Delta S_{10} 
\\ \nonumber
  \Delta S_{20} \, \mbox{Tr\,} M_1 
+ \Delta S_{10} \, \mbox{Tr\,} M_2 &=& 2 \Delta S_{20} + 2 \Delta S_{10} \; ,
\end{eqnarray}
where $\Delta S_{ij} = (S_i - S_j)/2$ and $T_0 = T_1 = T_2$.
The index 0 denotes the central orbit and the indices
1 and 2 the two satellite orbits. Only two of the three
equations in (\ref{sec2f}) are independent.
The relations (\ref{sec2f}) follow from the
normal form (\ref{sec2e}) [cf.\ equations (\ref{ap7}) and (\ref{ap9}) in
appendix \ref{seca1}]. For a general system they are only
valid in the vicinity of
the bifurcation. With increasing $\varepsilon$ they lose
their validity, and the transitional approximation
gradually becomes inaccurate. In order to obtain a formula which uniformly
interpolates over the region from the bifurcation up to regimes
where Gutzwiller's approximation is valid (without restrictions
on the semiclassical amplitudes) one has to consider
two modifications. First, higher-order corrections to
the normal form (\ref{sec2e}) cannot be neglected any more.
However, one can apply a mapping which brings the exponent
in (\ref{sec2b}) back to the normal form. Secondly, one has
to take into account the differences of the values of the
exponential prefactor in (\ref{sec2b}) at the different
stationary points. These steps are carried out
in appendix \ref{seca1}. In the following we discuss the
uniform approximation which is obtained there.

The uniform approximation for the joint semiclassical
contribution of orbits which are involved in a
generic period-quadrupling bifurcation is given by
\begin{equation} \label{sec3a}
d_\xi(E) \approx
\frac{1}{4 l \pi \hbar^2} \mbox{\bf Re}
\int_0^\infty \! \mbox{d}I \, 
[ T_0 + \alpha_1 I + \alpha_2 I^2 ] \,
\mbox{J}_0 \left( \frac{\tilde{b} I^2}{\hbar} \right) \,
\exp \left[ \frac{i}{\hbar} (S_0 - \tilde{\varepsilon} I 
- \tilde{a} I^2) - \frac{i \pi}{2} \nu \right] \; ,
\end{equation}
where $\mbox{J}$ denotes the Bessel function of the first kind and
\begin{eqnarray} \label{sec3b}
\tilde{\varepsilon} &=&
\frac{\sigma_{\tilde{\varepsilon}} T_0}{4 l A_0}
\; , \hspace{2cm}
\tilde{a} = \frac{\tilde{\varepsilon}^2 (\Delta S_{10}+\Delta S_{20})}{
16 \Delta S_{10} \, \Delta S_{20}}
\; , \hspace{2cm} 
\tilde{b} = \frac{\tilde{\varepsilon}^2 \Delta S_{21}}{
16 \Delta S_{10} \, \Delta S_{20}}
\nonumber \\
\alpha_1 &=& l \sigma_{\tilde{\varepsilon}} \tilde{\varepsilon}^2 \left[
\frac{(\Delta S_{10}+\Delta S_{20}) \, A_0}{\Delta S_{10} \, \Delta S_{20}}
- \frac{\Delta S_{20} \, A_1}{4\Delta S_{10} \, \Delta S_{21}}
\sqrt{\left| \frac{4\Delta S_{21}}{\Delta S_{20}} \right|}
- \frac{\Delta S_{10} \, A_2}{4\Delta S_{20} \, \Delta S_{12}}
\sqrt{\left| \frac{4\Delta S_{12}}{\Delta S_{10}} \right|} \right]
\nonumber \\
\alpha_2 &=& \frac{l |\tilde{\varepsilon}|^3}{4} \left[
\frac{A_0}{\Delta S_{10} \, \Delta S_{20}}
+ \frac{A_1}{4\Delta S_{12} \, \Delta S_{10}}
\sqrt{\left| \frac{4\Delta S_{21}}{\Delta S_{20}} \right|}
+ \frac{A_2}{4\Delta S_{21} \, \Delta S_{20}}
\sqrt{\left| \frac{4\Delta S_{12}}{\Delta S_{10}} \right|} \right] \; .
\end{eqnarray}
Equation (\ref{sec3a}) with definitions (\ref{sec3b}) 
is invariant under exchange of the indices 1 and 2.
The index $\nu$ and $\sigma_{\tilde{\varepsilon}}$,
the sign of $\tilde{\varepsilon}$, can be
determined from the Maslov indices of the periodic orbits.
The Maslov index of an unstable real satellite orbit is
always $\nu$, that of a stable real satellite orbit is
always $\nu - \sigma_{\tilde{\varepsilon}}$ and that of the
central orbit is $\nu + \sigma_{\tilde{\varepsilon}}$.
The actions of the real orbits are ordered in the same
way as their Maslov indices, i.\,e.\ $S_i > S_j$ if and
only if $\nu_i > \nu_j$. If both satellite orbits are
complex the sign of $\tilde{\varepsilon}$ is given by
$\sigma_{\tilde{\varepsilon}} = \mbox{sign}(S_1 - S_0) 
=\mbox{sign}(S_2 - S_0)$.

The coefficients (\ref{sec3b}) 
depend only on quantities which enter also the individual
contribution
(\ref{sec2e}). As an important consequence,
the joint contribution (\ref{sec3a}) is still invariant under canonical
transformations.
The transitional approximation is obtained if one
keeps only the first of the three pre-exponential terms in the
integrand.

There are some special cases of values of $\tilde{a}$ and
$\tilde{b}$ for which the integral (\ref{sec3a}) can be
evaluated analytically. They are discussed in appendix \ref{seca2}.
Numerically useful expressions that can be applied for arbitrary
values of $a$ and $b$ are given in appendix \ref{seca3}.
In the following we discuss different limits of the
integral (\ref{sec3a}). If the action differences 
$\Delta S_{ij}$ are large in comparison to $\hbar$
then a replacement of the Bessel function by its 
leading asymptotic term and a stationary-phase
evaluation of the integral yields a sum over 
Gutzwiller contributions (\ref{sec2c}) for the real
satellite orbits.
Complex satellites do not contribute since they cannot be reached by a
steepest-descent deformation of the integration manifold.
(A detailed study of contributions of complex orbits near
bifurcations can be found in \cite{Mou96}).
This is sensible since the complex satellites, though having complex
coordinates, still have real actions and semiclassical amplitudes
[cf.\ equations (\ref{ap7}) and (\ref{ap9})], and
their isolated contributions would not be exponentially suppressed
with $\hbar\to 0$. (In this respect, the role of the complex orbits
resembles that of the complex
satellite in a period-doubling bifurcation \cite{SS97}.)

The contribution of the central orbit
is given by the leading semiclassical contribution
from the boundary of the integral at $I=0$.

In the opposite limit $\tilde{\varepsilon} = 0$,
i.\,e.\ at the bifurcation, all action differences vanish.
The leading-order semiclassical
contribution of equation (\ref{sec3a}) is then given by
\begin{equation} \label{sec3c}
d_\xi(E) \approx \frac{T_0}{4 l \sqrt{2 \pi^3 \hbar^3 |b|}}
\mbox{\bf Re\,} \left\{ \left[ 
K \left( \sqrt{\frac{|b|+a}{2|b|}} \right) e^{-i\pi/4} +
K \left( \sqrt{\frac{|b|-a}{2|b|}} \right) e^{ i\pi/4} \right]
\exp \left( \frac{i}{\hbar} S_0 - \frac{i \pi}{2} \nu \right)
\right\} \; ,
\end{equation}
if $|a|<|b|$, and by
\begin{equation} \label{sec3d}
d_\xi(E) \approx \frac{T_0}{4 l \sqrt{\pi^3 \hbar^3 (|a|+|b|)}} 
K \left( \sqrt{\frac{2|b|}{|a|+|b|}} \right) \,
\cos \left( \frac{S_0}{\hbar} - \frac{\pi}{2} \nu 
-\frac{\pi}{4} \sigma_{a} \right) \; ,
\end{equation}
if $|a|>|b|$. $K(z)$ denotes the complete elliptic integral of 
the first kind. 
$T_0$ and $S_0$ are, respectively, the period and action
of the orbits at the bifurcation,
$a$ and $b$ are the coefficients in the normal form (\ref{sec2e})
for $\varepsilon = 0$, and $l=r/4$ is the repetition number of the
satellite orbits. The contributions (\ref{sec3c}) and (\ref{sec3d})
are by an order $\hbar^{-1/2}$ larger than the contribution of an
isolated period orbit, i.\,e.\ the singularity index
of the bifurcation is $1/2$. Although it is not written
explicitly, all quantities in (\ref{sec3c})
and (\ref{sec3d}) depend on the integer $l$.
In detail, $T_{0,l} = l T_{0,l=1}$, $S_{0,l} = l S_{0,l=1}$, 
$\nu_l = l \nu_{l=1}$ and $a_l = l a_{l=1}$ and $b_l = l b_{l=1}$.
It follows that the amplitude of the contribution at the bifurcation
decreases like $l^{-1/2}$ with increasing $l$. However, it cannot be
expected that this approximation is good for arbitrarily large $l$.
For longer periodic orbits bifurcations tend to occur more frequently.
It is expected that for larger $l$ there are other bifurcations which
interfere with the considered bifurcation.

The formulas of this section can also be applied, with minor
modifications, to two-dimensio\-nal area-preserving maps
whose time-evolution is governed by the Floquet operator $F$.
These maps correspond to systems with one degree of freedom
whose Hamiltonian operator is
periodic in time, $H(t+T) = H(t)$, and $F = U(T)$ is the unitary
time-evolution operator for one period. The Floquet operator
has unimodular eigenvalues $e^{-i\phi_i}$ with phases
$\phi_i$ that are called quasi-energies. The
quasi-energies can be determined from a knowledge of 
the traces of powers of $F$.

For maps the trace formula approximates $\Tr F^n$ semiclassically
instead of the level density. Furthermore, one has to pay attention
to the following differences in comparison to
autonomous systems with two degrees of freedom:
(i) The orbits which contribute are those with a fixed
period $n$, not those with a given energy $E$;  
(ii) the primitive periods have to be expressed in units
of $T$ and thus are integer valued;
(iii) the action is not the reduced energy-dependent one,
but depends on time (that is, on the number $n$);
(iv) instead of taking twice the real part, the full complex
contribution has to be taken;
(v) the results further differ by a factor $2\pi\hbar$.
 
It follows that the contribution $C^{(n)}_\xi$ of a
period-quadrupling bifurcation to $\Tr F^n$ is given by
\begin{equation} \label{sec3e}
C^{(n)}_\xi \approx \frac{1}{4 l \hbar} \int_0^\infty \! \mbox{d}I \, 
[ n + \alpha_1 I + \alpha_2 I^2 ] \,
\mbox{J}_0 \left( \frac{\tilde{b} I^2}{\hbar} \right) \,
\exp \left\{ \frac{i}{\hbar} (S_0 - \tilde{\varepsilon} I 
- \tilde{a} I^2) - \frac{i \pi}{2} \nu \right\} \; ,
\end{equation}
where the quantities $\tilde{a}$, $\tilde{b}$,
$\tilde{\varepsilon}$, $\alpha_1$ and $\alpha_2$
are determined by (\ref{sec3b}) with $T_0 = n$
and $A_i = n/(r_i \sqrt{|\Tr M_i - 2|})$.
Here $r_i$ and $M_i$ are the repetition number and
monodromy matrix of the orbit, respectively.

\section{Numerical results}

We now test the uniform approximation numerically on the example
of a periodically kicked top \cite{HKS87,KHE93,BGHS96} and
compare the results to those for the transitional approximation (valid
close to the bifurcation) and
the Gutzwiller approximation (which treats the orbits as being
isolated). The kicked top is a dynamical system that involves
the angular-momentum operators $J_x$, $J_y$, $J_z$ which
satisfy the usual commutation relations $[J_k,J_l]=i\epsilon_{klm}J_m$,
where $\hbar$ is set to unity. The evolution of the system
conserves the total angular momentum $J_x^2+J_y^2+J_z^2=j(j+1)$.
This introduces the quantum number $j$ which fixes the dimension
$2j+1$ of the Hilbert space. $j+1/2$ further plays the role of
the inverse of Planck's constant, and the semiclassical limit
is reached by $j\to\infty$. After normalization of the angular-momentum
vector the phase space of the classical system is
revealed as the unit sphere.

The specific top that is considered here is described by
the Floquet operator
\begin{equation}\label{sec4a}
F=\exp\left[-i\frac {k_z}{2j+1}J_z^2-ip_zJ_z\right]\exp\left[-ip_yJ_y\right]
\exp\left[-i\frac {k_x}{2j+1}J_x^2-ip_xJ_x\right] \; .
\end{equation}
This describes rotations by angles $p_i$ and nonlinear
rotations (torsions) of strength $k_i$. For the study of
bifurcations we hold the values of the $p_i$ fixed ($p_x=0.3$,
$p_y=1.0$, $p_z=0.8$) and vary $k=k_z=10k_x$ as a control parameter.
The classical counterpart of the system
is integrable for $k=0$ and displays well developed chaos 
at $k=5$.

At $k=0$ the top describes a linear rotation. In this situation the
system has only two periodic orbits, both of period one and positioned at the
intersection of the rotation axis with the spherical phase space.
They are called the trivial periodic orbits.
As $k$ is increased, new orbits show up in bifurcations. 
The first two period-quadrupling bifurcations are encountered
at $k=k^{(1)} = 1.0055\ldots$ and
$k=k^{(2)} = 1.1954\ldots$\,\,. Both have one of the trivial orbits in their
center. The next period-quadrupling happens at
$k=k^{(3)} = 3.0336\ldots$\,\,. The central orbit of this
bifurcation
is born at a smaller value of $k=k^{(4)}=2.4497\ldots$
in a tangent bifurcation together with an unstable partner. 
The form of the three period-quadrupling
bifurcations is of type $|a| > |b|$, i.\,e.\ on
one side of the bifurcation ($k < k^{(1,2,3)}$) both satellite
orbits are complex and on the other side ($k > k^{(1,2,3)}$)
both satellites are real.

Contributions of periodic orbits engaged in period-quadrupling
bifurcations first show up in the trace $\Tr F^4$ since the
repetition number of the central orbit has to be a multiple of four.
In the following we describe different semiclassical contributions
to $\Tr F^4$.
In general, orbits of primitive period one, two, and four
enter this trace. Complex orbits with the same primitive periods
can also contribute, however they
must be reachable by a steepest-descent contour deformation.
All these orbits can participate in bifurcations. 
One has to deal with tangent bifurcations of orbits of primitive
period one, two, and four, period-doubling bifurcations with central
orbits of primitive period one and two, and period-quadrupling
bifurcations with central orbits of primitive period one.
For $0\leq k \lesssim 2.5$, however, only few orbits are relevant, namely,
those which participate in the period-quadrupling bifurcations at $k^{(1)}$
and $k^{(2)}$, and in the tangent bifurcation at $k^{(4)}$.

\begin{figure}[htbp]
\begin{center}
\mbox{\epsfxsize17cm\epsfbox{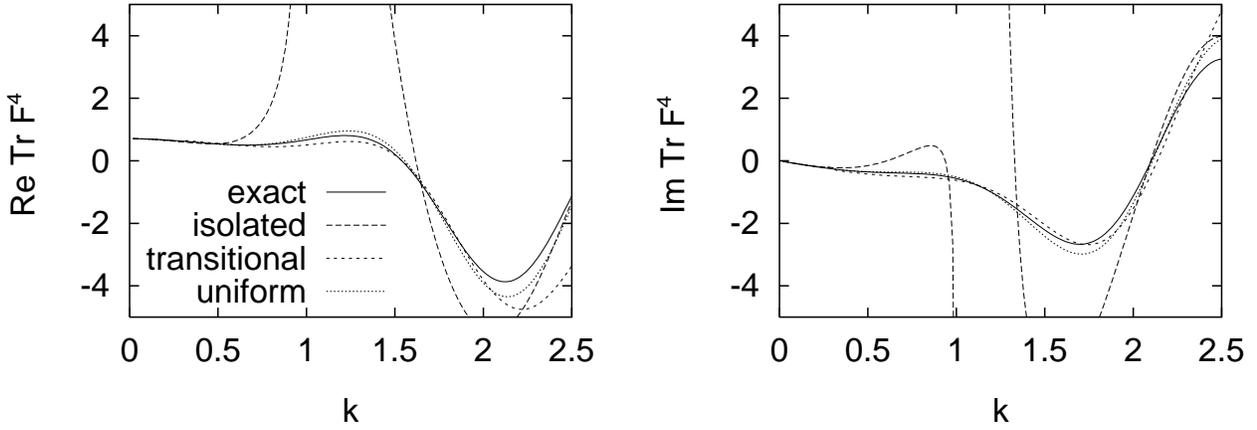}}
\end{center}
\caption{The trace $\Tr F^4$ of the 4th power of the Floquet operator
(\protect\ref{sec4a}) as a function
of the control parameter $k$ for $j=4$.
The plots show the exact quantum result, the
uniform approximation, the transitional approximation,
and the approximation which considers the orbits as
isolated.
One observes that the isolated approximation 
diverges at the period-quadrupling bifurcations while the
transitional and the uniform approximation behave regularly there.
However, the transitional approximation starts to break down
far to the right of the bifurcations.
}
\label{fig:Trk}
\end{figure}

\begin{figure}[htbp]
\begin{center}
\mbox{\epsfxsize17cm\epsfbox{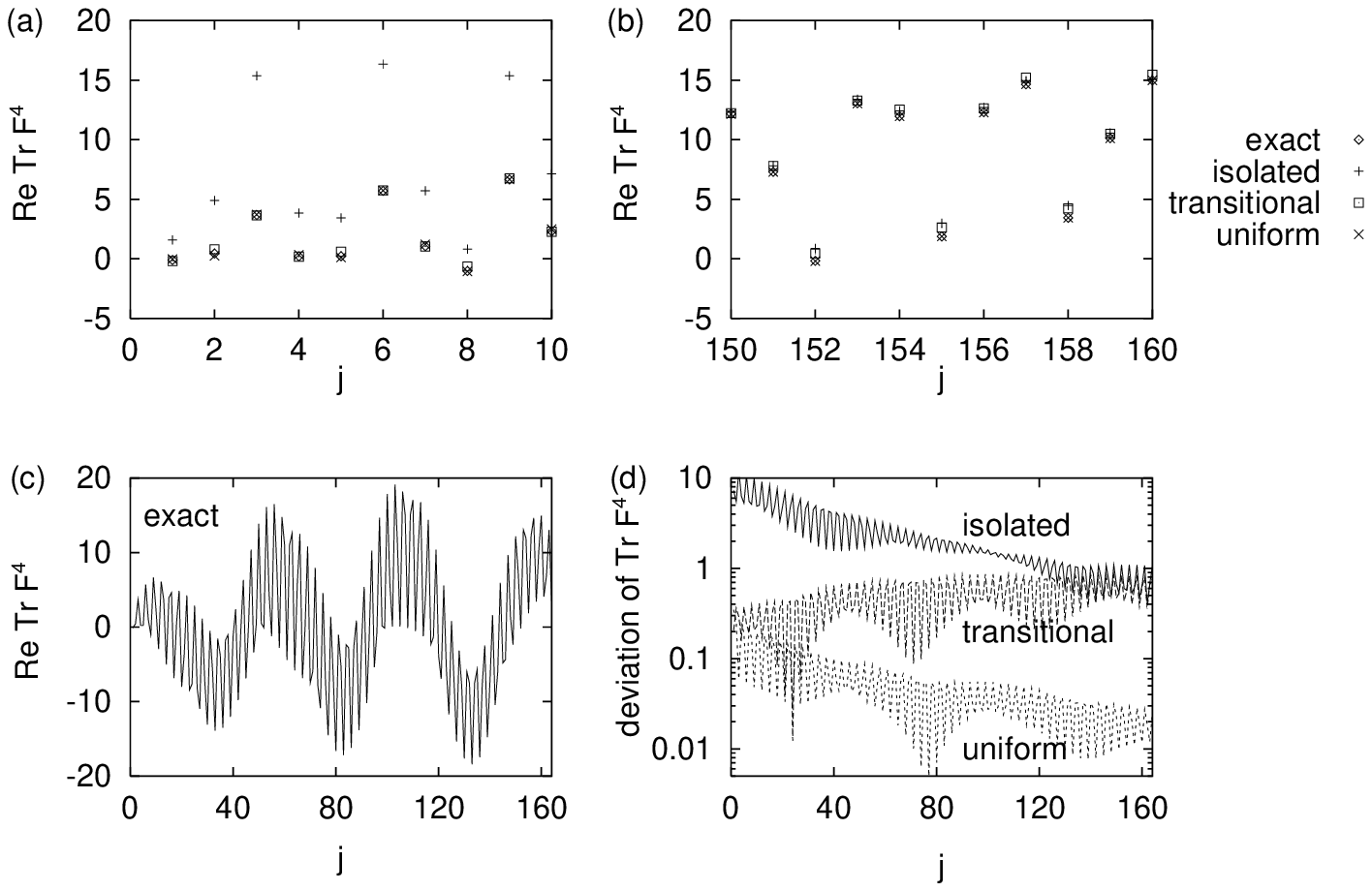}}
\end{center}
\caption{The trace $\Tr F^4$
as a function of  $j$ for $k=1.5$.
(a,b)
The real part of the exact quantum result, the
uniform approximation, the transitional approximation,
and the approximation which considers the orbits as
isolated.
(c) Illustration of the oscillations of the exact trace.
(d)
The error of the semiclassical traces,
measured by the absolute value of their deviation from the exact one.
}
\label{fig:Trj}
\end{figure}

In figure~\ref{fig:Trk} the trace $\Tr F^4$ for $j=4$ is plotted
against $k$. The exact result is compared
to three semiclassical approximations
which, respectively, treat the two period-quadrupling bifurcations
by the uniform approximation, the transitional approximation,
and the approximation that considers the orbits as isolated.
The tangent bifurcation is described in all three cases by
the uniform approximation of \cite{SS97}.
One observes that the isolated-orbit approximation diverges at the
period-quadrupling bifurcations whereas the transitional and the
uniform approximation behave regularly there. The transitional
approximation, however, loses accuracy for the largest $k$-values
in the displayed range, which is most clearly seen in the real
part of the trace.

In figure~\ref{fig:Trj} we investigate the behaviour of the
semiclassical approximations as the semiclassical limit is
approached, i.\,e.\ we fix the parameter $k$ at $k=1.5$
and increase the value of $j$.
For low values of $j$ the isolated approximation 
shows large deviations while the transitional and the uniform approximation 
are accurate. For larger values of $j$ the sum of isolated
contributions gains validity, since the effective Planck's constant
$1/(j+1/2)$ becomes small in comparison to the action differences
of the orbits. The transitional approximation on the other hand becomes
slightly more inaccurate, since the error in the semiclassical amplitudes
shows up more strongly when the orbits can be considered as isolated.
Figure~\ref{fig:Trj}(d) shows
the deviation $|\Tr F^4_{\rm sc}-\Tr F^4_{\rm qm}|$ of the semiclassical traces
$\Tr F^4_{\rm sc}$ from the exact trace $\Tr F^4_{\rm qm}$. 
This function reveals a crossover between 
the transitional and the isolated approximation and displays the
superiority of the uniform approximation over the whole range of $j$:
The uniform approximation is up to an order of magnitude
more accurate than the other two approximations.

\begin{figure}[htbp]
\begin{center}
\mbox{\epsfxsize12.5cm\epsfbox{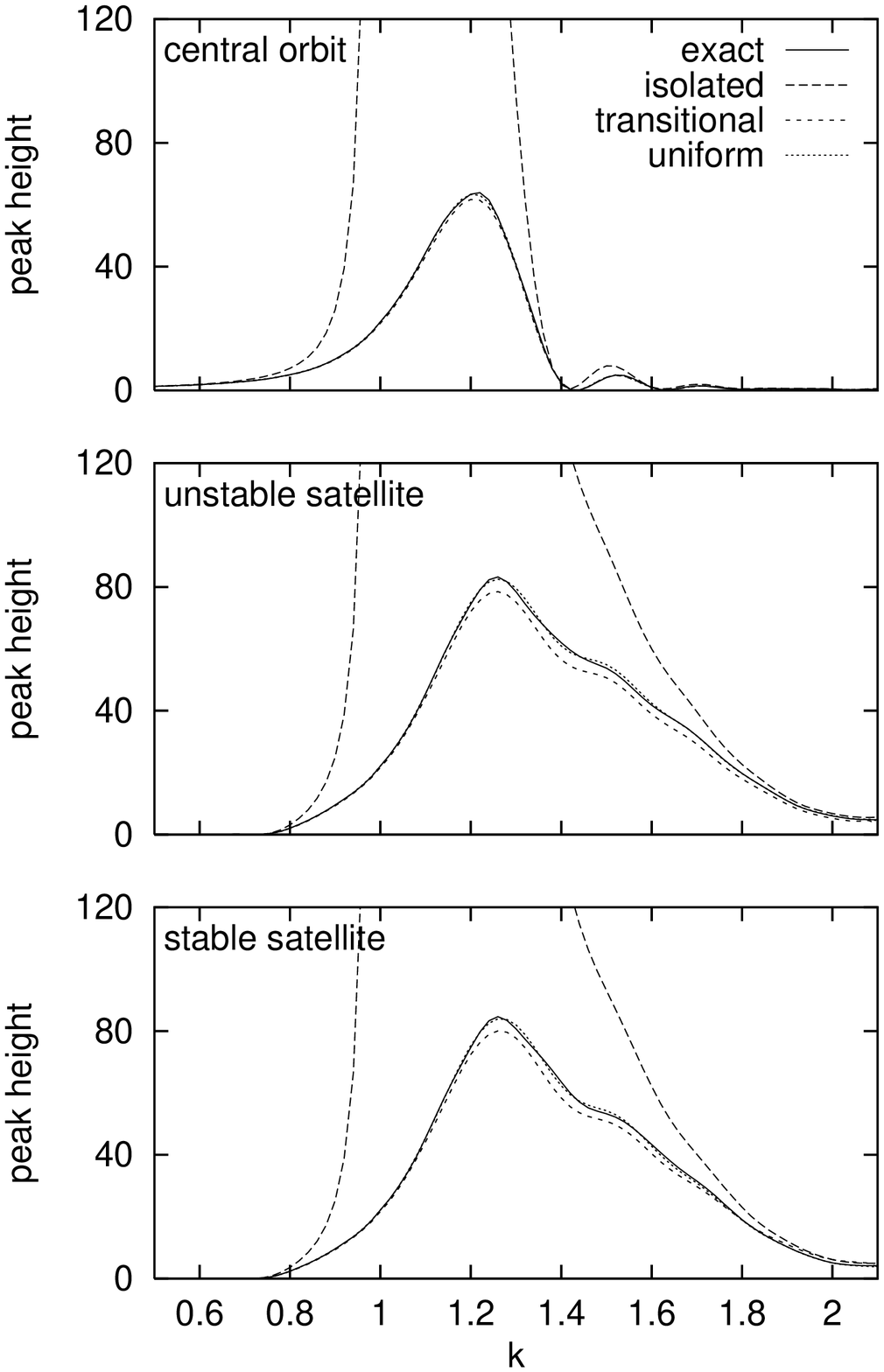}}
\end{center}
\caption{Peak heights $|T^{(4)}(S_{\rm cl}(k))|^2$ at the
values of the actions of three periodic orbits as a function
of the control parameter $k$. As $k$ goes through $k^{(1)} =
1.0055\ldots$ the three orbits coalesce in a generic period-quadrupling
bifurcation.
The plots show the exact quantum result, the
uniform approximation, the transitional approximation
and the approximation which considers the orbits as
isolated. They are evaluated with $j_{\rm min}=1$ and $j_{\rm max}=64$.
}
\label{fig:Tk}
\end{figure}

Figure~\ref{fig:Trj}(c) illustrates an oscillatory behaviour of $\Tr F^4$
which semiclassically originates from the interference of the
contributions from the two period-quadrupling bifurcations.
These contributions can be separated by considering the function
\begin{equation} \label{four}
T^{(n)}(S)=\frac 1 {j_{\rm max}-j_{\rm min}+1}
\sum_{j=j_{\rm min}}^{j_{\rm max}}e^{-ijS}\mbox{Tr}\,F^n(j) \;,
\end{equation}
which has peaks at the positions of the actions of the
periodic orbits.
In its essence this function is a Fourier coefficient
of $\mbox{Tr}\,F^n$ with respect to $j$.
It allows one to study the contributions of
periodic-orbit clusters individually in case that the action
differences of orbits from different clusters are sufficiently large.
A convenient testing tool then is the study of the peak height $T^{(n)}(S)$
at the value of the classical action $S_{\rm cl}$ of a given orbit in a 
cluster as a function of $k$ or for different values of $j_{\rm min}$
and $j_{\rm max}$ in (\ref{four}).

In figure~\ref{fig:Tk} we show a quantum-mechanical evaluation of
$|T^{(4)}(S_{\rm cl})|^2$ for the three orbits involved in the
period-quadrupling at $k^{(1)}$
as the parameter $k$ is steered across the bifurcation.
We use $j_{\rm min}=1$ and $j_{\rm max}=64$.
The exact quantum-mechanical
curves are compared to results for the uniform,
the transitional, and the isolated approximation.
The uniform approximation
is excellent over the whole range of $k$. It can hardly be
distinguished from the quantum result. The approximation
in terms of isolated orbits on the other hand fails completely.
It diverges at the bifurcation and only gains validity
again for values of $k$ where the amplitudes of the contributions
are already quite small. The transitional approximation is good
at the bifurcation; however, as $k$ is increased a clear deviation
from the quantum result can be seen for the satellite orbits.
This result shows again that there is a region where the uniform
approximation is essential, since both other approximations,
the transitional approximation and the approximation in terms of
isolated orbits fail or are inaccurate.

\begin{figure}[htbp]
\begin{center}
\mbox{\epsfxsize17cm\epsfbox{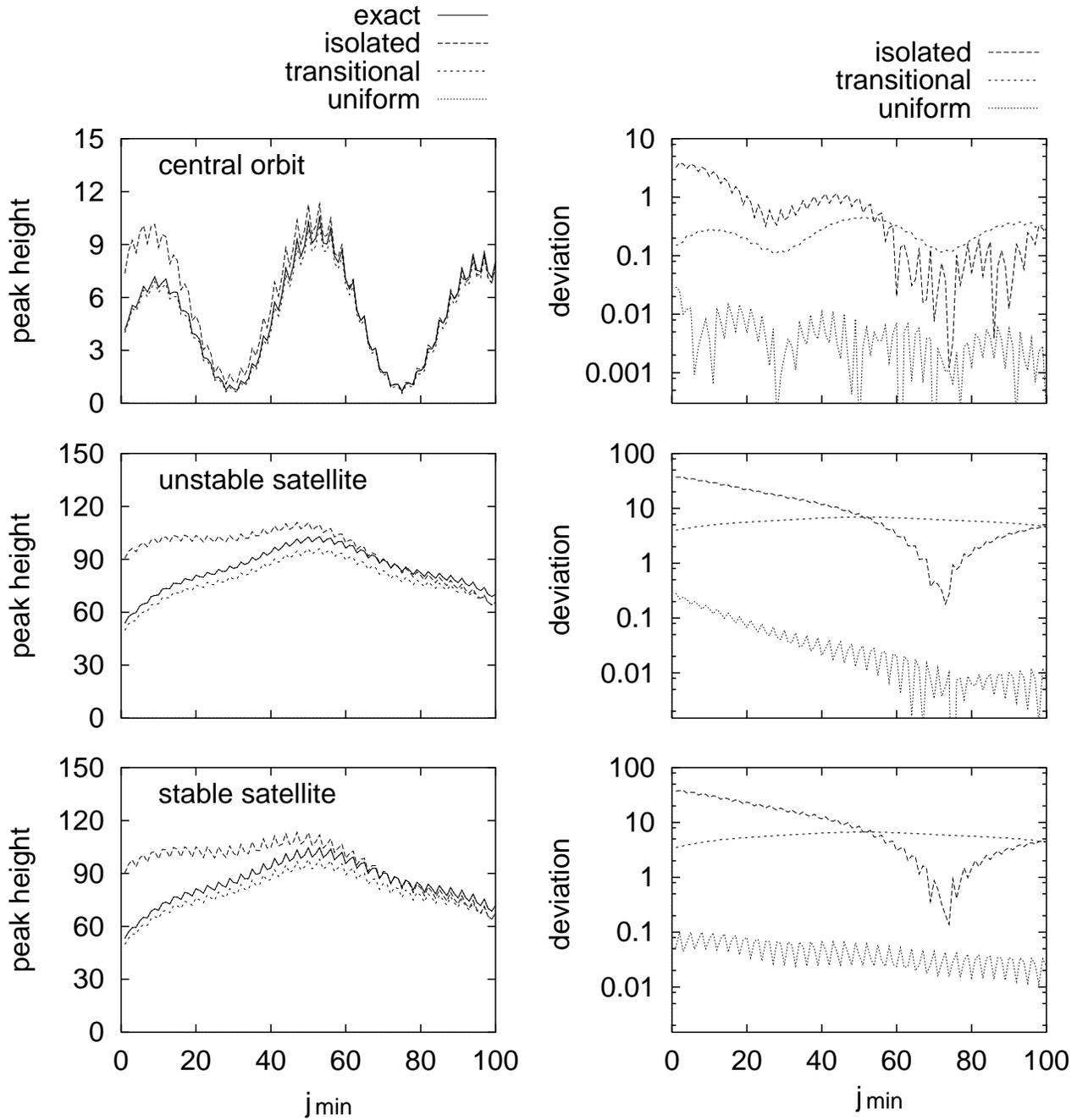}}
\end{center}
\caption{Peak heights $|T^{(4)}(S_{\rm cl})|^2$ at the
values of the actions of three periodic orbits as a function
of $j_{\rm min}$ with $j_{\rm max}=j_{\rm min}+63$ and $k=1.5$.
The plots on the left show the exact quantum result, the
uniform approximation, the transitional approximation,
and the approximation which considers the orbits as
isolated. The plots on the right show the deviation between
the semiclassical and the exact peak heights.
}
\label{fig:Tj}
\end{figure}

One can also investigate the peak
heights for a given $k$ and increasing
$j_{\rm min}$ and $j_{\rm max}$. In figure~\ref{fig:Tj} 
the result is shown as a function of  $j_{\rm min}$ 
with $j_{\rm max}=j_{\rm min}+63$. Again, $k$ is set to $1.5$.
Deviations are visible for the transitional and the isolated
approximation. Once more a crossover in the accuracies of these
approximations is observed. 
The uniform approximation  is by far superior over the whole range
of $j_{\rm min}$
and cannot be distinguished from the exact result in the plots for the
peak heights. 

Our numerical studies reveal that
the approximation with isolated contributions
is valid only far away from bifurcations or for sufficiently small
values of Planck's constant
while the transitional
approximation does not gain validity in the semiclassical
limit since it doesn't involve
the correct semiclassical amplitudes. The uniform
approximation however gives 
reliable predictions
both close to a bifurcation as well as far away from it,
and its accuracy increases in the semiclassical limit. 

\section{Conclusions}

In the present paper we derived a uniform approximation
for the joint contribution of periodic orbits that
are involved in a generic period-quadrupling bifurcation. Together
with the results of \cite{Sie96,SS97} this completes
the uniform treatment of generic bifurcations in
autonomous systems with two degrees of freedom and two-dimensional
area-preserving maps. Generic here means that these
bifurcations are typically encountered in mixed
systems without symmetries as one parameter
of the system is varied. They are also called bifurcations
of codimension one since only one parameter needs to be changed
in order to bring the participating orbits into coalescence.
In systems with symmetries there can be additional kinds of
bifurcations. Some of them can be described by small
modifications of the formulas for generic bifurcations
\cite{SS97}.

In the following we discuss the limits of validity of the
uniform approximations and possible extensions.
Although in a generic situation one does not encounter
any other form of coalescence of periodic orbits than
those discussed above, there are still cases in which
the uniform approximations for generic bifurcations
have to be modified. The reason for this is that periodic
orbits can undergo several subsequent bifurcations.
One observes for instance that
the iteration of the map corresponding to a normal form
describes more periodic orbits of longer periods
and allows for additional bifurcations.
Another source of additional periodic orbits and bifurcations
are higher-order terms in extended normal forms.
If the bifurcations occur rapidly one after the other
they cannot be considered separately, and instead
all participating orbits have to be treated collectively.
The next step beyond the isolated treatment of a bifurcation
is the collective treatment of two subsequent bifurcations.
Often the
two consecutive bifurcations can
also be considered as being part of a bifurcation of codimension
two, since the two bifurcations can be
brought into coalescence by varying a second parameter.
A collective treatment of two subsequent bifurcations
then is necessary if one is sufficiently close to a
bifurcation of codimension two in parameter space.
The methods for obtaining uniform approximations for
these cases are, in principle, the same as for bifurcations
of codimension one; however, the normal forms are more
complicated. Examples for these normal forms are given
in \cite{SSD95,SD96}, and uniform approximations for
bifurcations of codimension two can be found in
\cite{Sch97,BRS97}. It is even possible to completely
semiclassically quantize a kicked top with low-dimensional
Hilbert space by including uniform approximations for
bifurcations of codimension two \cite{SH97}. 

As longer and longer periodic orbits are considered,
bifurcations tend to occur more rapidly
and it is expected that then bifurcations of
even higher codimensions become important.
This makes the semiclassical treatment of mixed
systems more and more complicated. Applications
of semiclassical and uniform approximations in mixed
systems are therefore most useful in cases where mainly
the shortest periodic orbits of a system are needed,
for example if one is interested in long-range 
fluctuations in a spectrum or if the contributions
of long periodic orbits are suppressed. Applications
of this kind are discussed e.\,g.\ in \cite{BB97}.

\bigskip \bigskip \noindent {\large \bf Acknowledgments} 
\vspace{2.5mm}

M.\ Sieber wishes to acknowledge financial support
by the Deutsche Forschungsgemeinschaft under contract 
No.\ DFG-Ste 241/6-1 and /7-2. 
H.\ Schomerus gratefully acknowledges support by the
Sonderforschungsbereich `Unordnung und gro{\ss}e Fluktuationen' of the
Deutsche Forschungsgemeinschaft.

\appendix

\section{Derivation of the uniform approximation}
\label{seca1}

We follow in this section a method from catastrophe
theory for obtaining a uniform approximation for
an oscillatory integral with nearly coincident
stationary points. By this method the exponent
of an oscillating integrand is mapped onto a standard
normal form with the same structure of stationary
points. A description of the method and references
to previous work can be found in \cite{Con76,Ber76}.
A rigorous treatment including higher-order correction
terms is given in \cite{Dui74}.

The contributions of the periodic orbits to the level density
are contained in the integral (\ref{sec2b}) 
\begin{equation} \label{ap1}
d_\xi(E) \approx \frac{1}{2 \pi^2 \hbar^2 r} \, \mbox{\bf Re} 
\int_{-\infty}^\infty \! \mbox{d}q' \, 
\int_{-\infty}^\infty \! \mbox{d}p  \, \, 
g(q',p) \, \, \exp \left[ \frac{i}{\hbar} f(q', p) 
- \frac{i \pi}{2} \nu \right] \; ,
\end{equation}
where
\begin{equation} \label{ap2}
f(q',p) = \hat{S}(q', p, E) - q' p
\end{equation}
and
\begin{equation} \label{ap3}
g(q',p) = \frac{\partial \hat{S}}{\partial E} \,
\sqrt{\frac{\partial^2 \hat{S}}{\partial p \partial q'}} \; .
\end{equation}
The absolute sign of the mixed derivative of $\hat{S}$ 
inside the square root has been dropped since it is positive
near the bifurcation
(as can be seen from the normal form (\ref{sec2e})).
It changes its sign only at conjugate points where the
index $\nu$ changes as well. By writing it without
absolute sign the index $\nu$ can be kept constant and the
additional phase arises instead from the square root when
its argument becomes negative. The energy dependence of 
the functions $f$ and $g$ is not written explicitly.

We consider in the following the contributions of three
periodic orbits to (\ref{ap1}) that undergo a generic
period-quadrupling bifurcation as the energy or an
external parameter of the system is varied. We assume
that any other stationary points of the exponent in
(\ref{ap1}) which correspond to different periodic orbits
are well separated from the stationary points that
correspond to the periodic orbits which participate
in the bifurcation.
This means, for example, that the energy or other
parameters of the system have to be limited to ranges
in which the orbits do not participate in any further
bifurcation. Under these conditions a uniform approximation
is derived for the joint contribution of the three
periodic orbits.

Near the bifurcation the generating function 
$\hat{S}(q', p, E)$ in (\ref{ap2}) is approximately
given by the normal form (\ref{sec2e}). From this
normal form one can obtain properties of the periodic
orbits as is done in the following. First the stationary
points of (\ref{ap2}) have to be determined.
There is one stationary point at the origin which
corresponds to the central periodic orbit.
The other stationary points are determined conveniently
in terms of canonical polar coordinates $I,\Phi$ with
$p = \sqrt{2I} \cos\Phi$,
$q' = \sqrt{2I} \sin\Phi$. In terms of these
coordinates the normal form is given by
\begin{equation} \label{ap4}
f(q'(I,\Phi), p(I,\Phi)) = S_0 - \varepsilon I - a I^2 -
b I^2 \cos(4 \Phi) \; ,
\end{equation}
and the stationary points of $f$ are determined by the
equations
\begin{equation} \label{ap5}
0 = \sin(4 \Phi) \; , \; \; \; \; \; 
0 = -\varepsilon - 2[a + b \cos(4 \Phi)] I \; .
\end{equation}
There are altogether eight solutions for (\ref{ap5}), four
with $\cos(4 \Phi)=1$ corresponding to the satellite orbit
which is labelled by 1 in the following, and four with
$\cos(4 \Phi)=-1$ corresponding to the orbit labelled by 2.
The values of $I$ at the stationary points follow as
\begin{equation} \label{ap6}
I_{1,2} = - \frac{\varepsilon}{
2 (a + \sigma_{1,2} b)} \; ,
\end{equation}
where $\sigma_1=1$ and $\sigma_2=-1$. The satellite orbits
are real if $I_i$ is positive, i.\,e.\ for
$\sigma_{\varepsilon} = -\sigma_{c_i}$ where 
$c_i = a + \sigma_i b$ with $i \in \{1,2\}$, and we abbreviate
the sign of a quantity $x$ by $\sigma_x$.
For negative $I_i$ the coordinates $p$ and $q$ become complex.
The evaluation of (\ref{ap4}) at the stationary
points leads to the values of the actions of the
two satellite periodic orbits 
\begin{equation} \label{ap7}
S_{1,2} = S_0 + \frac{\varepsilon^2}{
4 (a + \sigma_{1,2} b)} \;.
\end{equation}
One can see from (\ref{ap7}) that the action difference
between any two of the periodic orbits increases proportional
to $\varepsilon^2$ for small $\varepsilon$.
The traces of the stability matrices can be determined from
\begin{equation} \label{ap8}
\mbox{Tr} M = \left( 
\frac{\partial^2 \hat{S}}{\partial p \, \partial q'} 
\right)^{-1} \, \left( 1 + 
\frac{\partial^2 \hat{S}}{\partial p \, \partial q'} 
\frac{\partial^2 \hat{S}}{\partial p \, \partial q'} -
\frac{\partial^2 \hat{S}}{\partial p^2} 
\frac{\partial^2 \hat{S}}{\partial q'^2} \right)  \; ,
\end{equation}
which has to be evaluated at the stationary points. It yields
\begin{equation} \label{ap9}
\Tr M_0 = 2 - \varepsilon^2 \; , \; \; \; \; \;
\Tr M_1 = 2 + \frac{8 b}{a + b}
\varepsilon^2 \; , \; \; \; \; \; 
\Tr M_2 = 2 - \frac{8 b}{a - b}
\varepsilon^2 \; .
\end{equation}
Both the actions (\ref{ap7}) and the traces of the
stability matrices (\ref{ap9}) are real quantities
even for orbits with complex coordinates.

Finally the Maslov indices of the orbits are determined. They
are given by $\nu+(n_n-n_p)/2$ where $n_n$ and $n_p$ are
the number of negative and positive eigenvalues of the
matrix of second derivatives of $f$, respectively. They
follow as
\begin{equation} \label{ap10}
\nu_0 = \nu + \sigma_{\varepsilon} \; , \; \; \; \; \;
\nu_{1,2} = \nu + \frac{1}{2} 
(\sigma_{c_{1,2}} - \sigma_{1,2} \sigma_{b}) \; .
\end{equation}
{}From (\ref{ap9}) it follows that for an unstable satellite
orbit $\sigma_{1,2} \sigma_{b} = \sigma_{c_{1,2}}$,
whereas for a stable satellite orbit
$\sigma_{1,2} \sigma_{b} = -\sigma_{c_{1,2}}$.
Using further the condition for real orbits
$\sigma_{c_{1,2}} = -\sigma_{\varepsilon}$
one obtains
\begin{equation} \label{ap11}
\nu_0 = \nu + \sigma_{\varepsilon} \; , \; \; \; \; \;
\nu_s = \nu - \sigma_{\varepsilon} \; , \; \; \; \; \;
\nu_u = \nu \; ,
\end{equation}
where the indices $u$ and $s$ denote an unstable and
stable real satellite orbit, respectively.

We continue now with the evaluation of the uniform 
approximation. The equations (\ref{ap7}) and (\ref{ap9}) 
entail the conditions (\ref{sec2f}) which hold for $\varepsilon\to 0$.
If the distance
to the bifurcation is increased (by changing the 
energy or a parameter of the system) higher-order
corrections to the normal form cannot be neglected
any more, and the dependence of $\hat{S}(q', p, E)$ on $q'$
and $p$ becomes more complicated. The main step in
the derivation of the uniform approximation consists
in the application of a coordinate transformation
which then brings the exponent in (\ref{ap1}) again
into the form of the normal form (inside a region
in which the stationary points are located)
\begin{equation} \label{ap12}
f(q',p) = F(Q',P) \; ,
\end{equation}
with
\begin{equation} \label{ap13}
F(Q',P) = S_0 - \frac{\ds \tilde{\varepsilon}}{\ds 2}(Q'^2 + P^2)
- \frac{\tilde{a}}{4} (Q'^4 + 2 P^2 Q'^2 + P^4) 
- \frac{\tilde{b}}{4} (Q'^4 - 6 P^2 Q'^2 + P^4) \; .
\end{equation}
The parameters of $F(Q',P)$ are chosen in such a way that
the mapping from $(q',p)$ to $(Q',P)$ is one-to-one in a
neighbourhood containing the stationary points. This
can be achieved by mapping the stationary points $(q_i',p_i)$
of the left-hand-side of (\ref{ap12}) onto the stationary points
$(Q_i',P_i)$ of the right-hand-side of (\ref{ap12}), which
leads to the following condition
\begin{equation} \label{ap14}
f(q_i',p_i) = F(Q_i',P_i) \; ,
\end{equation}
from which the parameters $\tilde{a}$ and $\tilde{b}$ can be
determined.

Condition (\ref{ap14}) is already fulfilled for the
stationary point at the origin which corresponds to the
central orbit. 
For the other stationary points the evaluation
of (\ref{ap14}) leads back to (\ref{ap7}) where
the parameters $\varepsilon$, $a$ and $b$ now carry
a tilde. Solving these equations for $\tilde{a}$ and
$\tilde{b}$ results in
\begin{equation} \label{ap15}
\tilde{a} = \frac{\tilde{\varepsilon}^2 (\Delta S_{10}
+\Delta S_{20})}{16 \Delta S_{10} \, \Delta S_{20}}
\; , \; \; \; 
\tilde{b} = \frac{\tilde{\varepsilon}^2 \Delta S_{21}}{
16 \Delta S_{10} \, \Delta S_{20}} \; ,
\end{equation}
where $\Delta S_{ij} = (S_i - S_j)/2$. Note that the third
parameter $\tilde{\varepsilon}$ of the mapping is not fixed. The
reason for this is that the form (\ref{ap13}) contains
one more parameter than is actually needed. By a simple
scaling transformation $Q' \rightarrow \lambda Q'$,
$P \rightarrow \lambda P$ one can change one of the
three parameters into $\pm 1$. In analogy to (\ref{ap9}) we
define $\tilde{\varepsilon}$ by 
\begin{equation}\label{ap15a}
\tilde{\varepsilon}^2 = 2 - \Tr M_0
\end{equation}
which measures the distance to the
bifurcation. The sign of $\tilde{\varepsilon}$ is the
same as the sign of $\varepsilon$ and can be determined
from the Maslov indices (\ref{ap11}) of the central
orbit and one real satellite orbit (if both satellite
orbits are complex then $\sigma_{\tilde{\varepsilon}}
= \sigma_{\tilde{a}}$).
The difference between the quantities with tilde and
without tilde is that those without tilde are obtained
from a Taylor expansion around the central orbit whereas
the quantities with tilde follow from the mapping.
For $\tilde{\varepsilon} \rightarrow 0$ the mapping (\ref{ap12})
approaches the identity transformation $(Q',P)=(q',p)$ and
the quantities with tilde approach the ones without tilde.

The mapping (\ref{ap12}) transforms the integral (\ref{ap1}) into 
\begin{equation} \label{ap16}
d_\xi(E) \approx \frac{1}{2 \pi^2 \hbar^2 r} \, \mbox{\bf Re} 
\int_{-\infty}^\infty \! \mbox{d}Q' \, 
\int_{-\infty}^\infty \! \mbox{d}P  \, \, 
G(Q',P) \, \, \exp \left[ \frac{i}{\hbar} F(Q',P) 
- \frac{i \pi}{2} \nu \right] \; ,
\end{equation}
where
\begin{equation} \label{ap17}
G(Q',P) = g(q',p) \,
\det \left( \frac{\partial (q',p)}{\partial (Q',P)} \right) \; ,
\end{equation}
and the determinant in (\ref{ap17}) is the Jacobian of the
transformation which will be denoted by $J(Q',P)$ in the
following. 

The uniform approximation is obtained by writing the
function $G(Q',P)$ in the following form
\begin{equation} \label{ap18}
G(Q',P) = \alpha_0 - \alpha_1
\frac{\partial F}{\partial \tilde{\varepsilon}}
- \alpha_2 \frac{\partial F}{\partial \tilde{a}}
- \alpha_3 \frac{\partial F}{\partial \tilde{b}}
+ H_1(Q',P) \frac{\partial F}{\partial Q'} 
+ H_2(Q',P) \frac{\partial F}{\partial P} \; .
\end{equation}
In order for this representation to be correct, the
constants $\alpha_0, \dots, \alpha_3$ have to be
determined such that the right-hand-side of (\ref{ap18})
has the correct values at the stationary points.
The last two terms in (\ref{ap18}) vanish at the
stationary points. They can be neglected since 
after inserting (\ref{ap18}) into (\ref{ap16}) 
they lead to terms which are of order $\hbar$ smaller
than the other terms as can be seen by an integration
by parts. Furthermore, the constant $\alpha_3$ can be
set equal to zero, since after inserting (\ref{ap18})
into (\ref{ap16}) the integral proportional to
$\alpha_3$ can be expressed in terms of the integrals
that are proportional to the other $\alpha_i$, as
can be shown by another integration by parts.

The remaining parameters $\alpha_{0,1,2}$ in (\ref{ap18}) are
obtained by an evaluation of (\ref{ap17}) at the
stationary points. For that purpose the value of the
Jacobian $J(Q',P)$ at the stationary points has to
be determined. This is done by differentiating (\ref{ap12})
twice. With the notation $(z_1,z_2):=(q',p)$ and
$(Z_1,Z_2):=(Q',P)$ this results in
\begin{equation} \label{ap19}
\left. \frac{\partial^2 F}{\partial Z_k \partial Z_l}
\right|_{\stackrel{Q'=Q'_i}{\scriptscriptstyle P=P_i}}
= \sum_{m,n=1}^2 
\frac{\partial z_m}{\partial Z_k}
\frac{\partial z_n}{\partial Z_l}
\left. \frac{\partial^2 f}{\partial z_m \partial z_n}
\right|_{\stackrel{q'=q_i'}{\scriptscriptstyle p=p_i}} \; ,
\end{equation}
from which the Jacobian $J(Q',P)$ follows as
\begin{equation} \label{ap20}
J(Q',P) = \sqrt{\frac{\det \left[ 
\frac{\partial^2 F}{\partial Z_i \partial Z_j} \right]}{
\det \left[ \frac{\partial^2 f}{\partial z_k \partial z_l}
\right]}} \; .
\end{equation}
By using (\ref{ap8}) this result can be expressed in the
form 
\begin{equation} \label{ap21}
\left[ J(Q',P) \, \sqrt{\frac{\partial^2 \hat{S}}{
\partial p \partial q'}} \, \right]_{\stackrel{q'=q_i'}{
\scriptscriptstyle p=p_i}}
= \sqrt{\frac{\Tr \tilde{M}_i - 2}{\Tr M_i - 2}} \; ,
\end{equation}
where we defined
\begin{equation} \label{ap22}
\Tr \tilde{M}_0 = 2 - \tilde{\varepsilon}^2 \; , \; \; \; \; \;
\Tr \tilde{M}_1 = 2 + \frac{8 \tilde{b}}{\tilde{a} + \tilde{b}}
\tilde{\varepsilon}^2 \; , \; \; \; \; \; 
\Tr \tilde{M}_2 = 2 - \frac{8 \tilde{b}}{\tilde{a} - \tilde{b}}
\tilde{\varepsilon}^2 \; .
\end{equation}
As before the quantities with tilde approach the ones without
tilde as $\tilde{\varepsilon} \rightarrow 0$. We continue
now with the determination of the $\alpha_i$. By evaluating
(\ref{ap17}) at the stationary points with (\ref{ap3}),
(\ref{ap18}) and (\ref{ap21}) one obtains the following
conditions
\begin{equation} \label{ap23}
\alpha_0 + \alpha_1 \tilde{I}_i + \alpha_2 \tilde{I}_i^2 =
T_i \, \sqrt{\frac{\Tr \tilde{M}_i - 2}{\Tr M_i - 2}} \; ,
\end{equation}
where $\tilde{I}_1$ and $\tilde{I}_2$ are given by (\ref{ap6})
if the parameters on the right-hand-side of (\ref{ap6}) are given
a tilde, and $\tilde{I}_0 = 0$. The constants $\alpha_0$,
$\alpha_1$ and $\alpha_2$ follow from (\ref{ap23}) 
and the definition (\ref{ap15a}) as
\begin{eqnarray} \label{ap24}
\alpha_0 &=& T_0
\\ \nonumber
\alpha_1 &=& \frac{4 \tilde{a}}{\tilde{\varepsilon}} \alpha_0
- \frac{(\tilde{a}+\tilde{b})^2\,T_1}{\tilde{b} \tilde{\varepsilon}}
\sqrt{\frac{8\tilde{b}\tilde{\varepsilon}^2}{
(\tilde{a}+\tilde{b})(\Tr M_1 - 2)}}
+ \frac{(\tilde{a}-\tilde{b})^2\,T_2}{\tilde{b}
\tilde{\varepsilon}}
   \sqrt{\frac{ - 8\tilde{b}\tilde{\varepsilon}^2}{
(\tilde{a}-\tilde{b})(\Tr M_2 - 2)}}
\\ \nonumber 
\alpha_2 &=& \frac{4 (\tilde{a}^2-\tilde{b}^2)}{
\tilde{\varepsilon}^2} \left[
\alpha_0 - \frac{(\tilde{a}+\tilde{b})\,T_1}{2\tilde{b}}
\sqrt{\frac{8\tilde{b}\tilde{\varepsilon}^2}{(\tilde{a}+\tilde{b})
(\Tr M_1 - 2)}} + \frac{(\tilde{a}-\tilde{b})\,T_2}{2\tilde{b}}
\sqrt{\frac{ - 8\tilde{b}\tilde{\varepsilon}^2}{(\tilde{a}-\tilde{b})
(\Tr M_2 - 2)}} \, \right] \; .
\end{eqnarray}
This completely specifies the uniform approximation.
The relations (\ref{sec3b}) for the $\alpha_i$ follow from
(\ref{ap24}) by using (\ref{ap15}). The integral representation
for the uniform approximation is obtained by changing the
integration variables in (\ref{ap16}) to canonical polar
coordinates $I,\Phi$ with $P = \sqrt{2I} \cos\Phi$,
$Q' = \sqrt{2I} \sin\Phi$ and performing the 
integration over $\Phi$. With the approximation for
$G(Q',P)$ that is discussed after (\ref{ap18}) this results
in the final expression (\ref{sec3a}).

\section{Special cases}
\label{seca2}

\subsection{The case $S_0 = (S_1 + S_2)/2$}

In case that the action of the central orbit is the
mean of the actions of the satellite orbits, the diffraction
integral in (\ref{sec3a}) can be evaluated analytically
since $\tilde{a}=0$. It is a special case of the form
of the bifurcation with $|\tilde{a}| < |\tilde{b}|$.
The first 
satellite orbit is real if  $\sigma_{\tilde{\varepsilon}} 
= - \sigma_{\tilde{b}}$ and the second satellite if
$\sigma_{\tilde{\varepsilon}} = \sigma_{\tilde{b}}$.
We define in the following $\Delta S = (S_1 - S_0)/2$.
The integral in (\ref{sec3a}) can be evaluated with the relation
\begin{equation} \label{spe1b}
\int_0^\infty \! \mbox{d}I \, \mbox{J}_0 (I^2) \, \exp (- i \eta I)
= \frac{\sqrt{-i} \pi \eta}{16}
H_{1/4}^{(1)} \left( e^{i \pi (\sigma_\eta + 1)} 
\frac{\eta^2}{8} \right) \,
H_{1/4}^{(1)} \left( e^{i \pi \sigma_\eta} 
\frac{\eta^2}{8} \right) \; ,
\end{equation}
and the first two derivatives of this equation with
respect to $\eta$. As before, $\sigma_\eta$ denotes the sign
of $\eta$. Altogether one obtains the following result for
the uniform approximation
\begin{eqnarray}
d_\xi(E) &\approx& \frac{|\Delta S|}{2 \hbar^2} \, \mbox{\bf Re\,}
\left\{ \left[ \left( \frac{A_0}{2} + \frac{A_1}{4 \sqrt{2}}
+ \frac{A_2}{4 \sqrt{2}} \right) B_1
- i \sigma_{\tilde{b}} \left( \frac{A_1}{4 \sqrt{2}}
- \frac{A_2}{4 \sqrt{2}} \right) B_2 \right. \right.
\nonumber \\ && \left. \left. 
+ \left( \frac{A_0}{2} - \frac{A_1}{4 \sqrt{2}}
- \frac{A_2}{4 \sqrt{2}} \right) B_3 \right]
\exp \left( \frac{i}{\hbar} S_0 - \frac{i \pi}{2} 
(\nu + \sigma_{\tilde{\varepsilon}}) \right) \right\} \; ,
\end{eqnarray}
where
\begin{eqnarray}
B_1 &=& H^{(+)}_{1/4} (z) \, H^{(-)}_{1/4} (z)
        + (1 - i \sigma_{\tilde{\varepsilon}})
        H^{(-\sigma_{\tilde{\varepsilon}})}_{1/4} (z) \, 
        H^{(-\sigma_{\tilde{\varepsilon}})}_{1/4} (z)
\nonumber \\
B_2 &=& H^{(+)}_{ 1/4} (z) \, H^{(-)}_{-3/4} (z) +
        H^{(+)}_{-3/4} (z) \, H^{(-)}_{ 1/4} (z)
        + 2 (1 - i \sigma_{\tilde{\varepsilon}})
        H^{(-\sigma_{\tilde{\varepsilon}})}_{ 1/4} (z) \, 
        H^{(-\sigma_{\tilde{\varepsilon}})}_{-3/4} (z)
\nonumber \\
B_3 &=& H^{(+)}_{-3/4} (z) \, H^{(-)}_{-3/4} (z)
        + (1 - i \sigma_{\tilde{\varepsilon}})
        H^{(-\sigma_{\tilde{\varepsilon}})}_{-3/4} (z) \, 
        H^{(-\sigma_{\tilde{\varepsilon}})}_{-3/4} (z)    \; ,
\end{eqnarray}
and $z = |\Delta S|/\hbar$. We have chosen the following notation
for the Hankel functions:
$H^{(+)}_\nu(z) = H^{(1)}_\nu(z)$ and
$H^{(-)}_\nu(z) = H^{(2)}_\nu(z)$.

Near the bifurcation the actions of the two satellite orbits
and the traces of the monodromy matrices are given by
\begin{equation} \label{spe1a}
S_1 = S_0 + \frac{\varepsilon^2}{4 b} \; , \; \; \; \; 
S_2 = S_0 - \frac{\varepsilon^2}{4 b} \; , \; \; \; \;
\mbox{Tr\,} M_{1,2} = 2 + 8 \varepsilon^2 \; ,
\end{equation}
and in the limit $\tilde{\varepsilon} \rightarrow 0$ the
contribution
\begin{equation}
d_\xi(E) \approx \frac{T_0}{16 l \pi^2 \hbar^{3/2} |b|^{1/2}} \,
\Gamma^2 \left( \frac{1}{4} \right) 
\cos \left( \frac{S_0}{\hbar} - \frac{\pi}{2} \nu \right)
\end{equation}
is obtained,
which agrees with (\ref{sec3c}) for $a=0$.

\subsection{The case $S_1 = (S_0 + S_2)/2$}

This is again a case where the integral in (\ref{sec3a})
can be evaluated analytically. This can be seen by
considering the previous expression for the uniform
approximation (\ref{ap16}) where the integration is
performed in terms of the $P$ and $Q'$ variables.
The constants $\tilde{a}$ and $\tilde{b}$ now satisfy
the relation $\tilde{a} = 3 \tilde{b}$, and the part
of the double integral in (\ref{ap16}) with constant
pre-exponential factor splits into a product of single
integrals that can be evaluated. The bifurcation is an
example for the case $|\tilde{a}| > |\tilde{b}|$. Both
satellite orbits are real if $\sigma_{\tilde{\varepsilon}}
= - \sigma_{\tilde{b}}$, and complex if 
$\sigma_{\tilde{\varepsilon}} = \sigma_{\tilde{b}}$, 
and we define again $\Delta S = (S_1 - S_0)/2$.
The part of the integral in (\ref{ap16}) with 
constant pre-exponential factor is given by
\begin{equation} \label{spe2b}
\int_{-\infty}^\infty \! \mbox{d}p  \, 
\int_{-\infty}^\infty \! \mbox{d}q' \,
\exp \left(- \frac{i \tilde{\varepsilon}}{2 \hbar} (p^2 + q'^2)
- \frac{i \tilde{b}}{\hbar} (p^4 + q'^4) \right) 
= \frac{2 \pi^2 |\Delta S| C_1^2}{|\tilde{\varepsilon}|} 
\exp \left( \frac{i}{\hbar} 2 \Delta S \right) \; ,
\end{equation}
where
\begin{equation}
C_1 = J_{-1/4} \left(\left|\frac{\Delta S}{\hbar}\right|\right)
e^{-i \sigma_{\tilde{b}} \pi/8} - \sigma_{\tilde{b}}
\sigma_{\tilde{\varepsilon}}
      J_{ 1/4} \left(\left|\frac{\Delta S}{\hbar}\right|\right)
e^{ i \sigma_{\tilde{b}} \pi/8} \; ,
\end{equation}
and $\sigma_{\tilde{b}} = \mbox{sign\,}(\tilde{b})
= \mbox{sign\,}(\Delta S)$.
The terms with $I$ or $I^2$ in the exponential prefactor
can be obtained from the first two derivatives of (\ref{spe2b})
with respect to $\tilde{\varepsilon}$.

We further define 
\begin{equation}
C_2 = J_{ 3/4} \left(\left|\frac{\Delta S}{\hbar}\right|\right)
e^{ i \sigma_{\tilde{b}} 3 \pi/8} - \sigma_{\tilde{b}}
\sigma_{\tilde{\varepsilon}}
      J_{-3/4} \left(\left|\frac{\Delta S}{\hbar}\right|\right)
e^{-i \sigma_{\tilde{b}} 3 \pi/8} \; ,
\end{equation}
and obtain for the uniform approximation (up to higher-order
corrections in $\hbar$)
\begin{eqnarray}
d_\xi(E) &\approx& \frac{|\Delta S|}{\hbar^2} \, \mbox{\bf Re\,}
\left\{ \left[ \left( \frac{A_0}{4} + \frac{A_1}{4 \sqrt{2}}
+ \frac{A_2}{8} \right) C_1^2
- \left( \frac{A_0}{2} - \frac{A_2}{4} \right) C_1 C_2 \right. \right.
\nonumber \\ && \left. \left. 
+ \left( \frac{A_0}{4} - \frac{A_1}{4 \sqrt{2}}
+ \frac{A_2}{8} \right) C_2^2 \right]
\exp \left( \frac{i}{\hbar} S_1 - \frac{i \pi}{2}
\nu \right) \right\} \; .
\end{eqnarray}

In the vicinity of the bifurcation the actions of the two satellite
orbits and the traces of the monodromy matrices are given by
\begin{equation} \label{spe2a}
S_1 = S_0 + \frac{\varepsilon^2}{16 b} \; , \; \; \; \; 
S_2 = S_0 + \frac{\varepsilon^2}{ 8 b} \; , \; \; \; \; 
\mbox{Tr\,} M_1 = 2 + 2 \varepsilon^2  \; , \; \; \; \; 
\mbox{Tr\,} M_2 = 2 - 4 \varepsilon^2  \; ,
\end{equation}
and in the limit $\varepsilon \rightarrow 0$ the following 
contribution is obtained,
\begin{equation}
d_\xi(E) \approx \frac{T_0}{32 l \pi^2 \hbar^{3/2} |b|^{1/2}} \,
\Gamma^2 \left( \frac{1}{4} \right) \,
\cos \left( \frac{S_0}{\hbar} - \frac{\pi}{2} \nu 
- \frac{\pi}{4} \sigma_{b} \right) \; .
\end{equation}
This agrees with (\ref{sec3d}) for $a=3b$.

\subsection{Further special cases}

We briefly discuss two additional cases that occur if 
$\tilde{b} = 0$ or $|\tilde{b}| = |\tilde{a}|$. 

In the case $\tilde{b} = 0$ the actions of the two satellite orbits
$S_1$ and $S_2$ are identical. This can occur in integrable
systems where both orbits are part of a torus. During the
bifurcation this whole torus of orbits arises. With an
integration by parts the formula (\ref{sec3a}) reduces 
in the limit $\tilde{b} = 0$ to 
\begin{equation} \label{b0}
d_\xi(E) \approx \frac{1}{4 l \pi \hbar^2} \mbox{\bf Re}
\int_0^\infty \! \mbox{d}I \, 
[ T_0 + \alpha I] \,
\exp \left\{ \frac{i}{\hbar} (S_0 - \tilde{\varepsilon} I 
- \tilde{a} I^2) - \frac{i \pi}{2} \nu \right\} \; ,
\end{equation}
and thus can be expressed by a Fresnel integral. The
constant $\alpha$ is given by
\begin{equation}
\alpha = \lim_{\tilde{b} \rightarrow 0} \left( \alpha_1 -
\frac{\tilde{\varepsilon}}{2 \tilde{a}} \alpha_2 \right) \; .
\end{equation}
A discussion of semiclassical approximations for bifurcations
in which a torus arises from a stable orbit is given in
\cite{Ric82}.

The other case $|\tilde{b}| = |\tilde{a}|$ separates the two forms
in which the generic period-quadrupling bifurcation can occur.
The set of stationary points in the normal form (\ref{ap13})
corresponding to one of the two satellite orbits goes to infinity
as $\tilde{b}$ approaches $\pm \tilde{a}$. This normal form is not
appropriate for a description of this case and correction terms
have to be added to it. 

\section{Series expansion of the uniform approximation}
\label{seca3}

The uniform approximation (\ref{sec3a}) consists of three 
integrals of type
\begin{equation}\label{seca3a}
I_\nu
\equiv \int_0^\infty x^\nu \exp\left[-i \gamma x-i\alpha x^2\right]
\mbox{J}_0 (x^2)\,\mbox{d}x
\end{equation}
with $\nu=0,1,2$. We now present power series 
in the coefficients $\alpha$ and $\gamma$ that are useful for
a numerical evaluation of these integrals.
In the case $|\alpha|>1$ (which corresponds to $|b/a|<1$ in the
normal form) one expands the integrand around $\gamma=0$
and uses the analytic continuation of the
integral 6.621.1 in \cite{GR94}. Expressing a
hypergeometric function by its defining series and applying the
duplication formula of the Gamma function one arrives at
\begin{equation}
I_\nu
=
\frac 12 \sum_{m=0}^\infty\sum_{n=0}^\infty
\frac{(\gamma/i)^n}{(i\alpha)^{\frac{n+\nu+1}2}}
\left(\frac{1}{2\alpha}\right)^{2m}
\frac{\Gamma\left(\frac{n+\nu+1}2+2m\right)}{n!(m!)^2}\;.
\end{equation}
For $|\alpha|<1$, i.\,e.\ $|b/a|>1$ in the normal form, 
one uses  the integrals 6.699.1 and 6.699.2 in \cite{GR94}. Formally,
a small imaginary part has to be added to the coefficients in order to
assure convergence. At the end this imaginary part is sent to zero.
After some transformations similar to those described above one arrives at
\begin{eqnarray}
I_\nu&=&\frac {\sqrt{2^\nu}} {\sqrt{8\pi}}
\sum_{n,m=0}^\infty\frac{(\sqrt 2\gamma/i)^n\alpha^{2m}}{m!n!}
\\
&&\times
\left[
\sin \left({\textstyle\frac{1+n+\nu}4\pi}\right)
\frac{\Gamma^2\left(\frac{1+n+\nu}4+m\right)}{\Gamma\left(\frac12+m\right)}
-i
\alpha
\cos \left({\textstyle\frac{1+n+\nu}4\pi}\right)
\frac{\Gamma^2\left(\frac{3+n+\nu}4+m\right)}{\Gamma\left(\frac32+m\right)}
\right]
\nonumber
\;.
\end{eqnarray}

\bibliographystyle{my_unsrt}
\bibliography{../Tex/paper}

\end{document}